\documentclass[11pt]{article}

\usepackage{amsmath}
\usepackage{amssymb}
\usepackage{graphicx}
\usepackage{booktabs}
\usepackage[section]{placeins}
\usepackage{flafter}
\usepackage{float}
\usepackage{xurl}
\usepackage{tabularx}
\usepackage{array}

\newcolumntype{L}{>{\raggedright\arraybackslash}X}
\newcolumntype{C}{>{\centering\arraybackslash}X}

\title{
Constrained Color Carrier:
Characterization-Preserving Conditional Color Rendering
in Multi-Illuminant Camera Profiles
}

\author{
Xilai Liang\\
\small Guangdong Technion--Israel Institute of Technology\\
\small Shantou, Guangdong, China
}

\date{}

\begin{document}

\maketitle

\begin{abstract}
In Digital Negative (DNG) multi-illuminant profiles, characterization matrices
and nonlinear rendering payloads share condition-dependent interpolation
slots, so adding a slot for rendering capacity also introduces an additional
characterization state. We introduce Constrained Color Carrier (CCC), which
constructs the three pre-serialization ColorMatrix and ForwardMatrix states
from the original dual matrix segments while allowing the same shared slots
to carry three HueSatMap rendering bases. Condition-specific HueSatMap
payloads are solved as numerical preimages of the target through a fixed
serialized downstream rendering operator, and carrier selection enforces a
preservation bound on the host-effective interpolated characterization. We
evaluate CCC on the Sony ILCE-7RM5 Adobe Standard dual-illuminant profile
using a white-balance-dependent Standard ColorCorrect target recovered from
Phocus 4.0.1 for the Hasselblad X2D 100C. Using Euclidean Oklab error, CCC
yields a worst-temperature 95th-percentile preservation error of 0.003857,
below the prescribed tolerance of 0.004, and reduces the worst-condition
95th-percentile target error from 0.08208 for the Dual representation to
0.03508. The synthetic Ordinary Triple yields a preservation error of
0.014802 and exceeds the same tolerance. The serialized DNG Camera Profile (DCP) and Extensible Metadata Platform (XMP) artifacts
reproduce the final color-table payloads exactly and yield zero numerical
difference from solver-side offline execution, showing that the CCC solution
is representable within the evaluated serialized profile format.
\end{abstract}

\section{Introduction}
\label{sec:introduction}

Digital camera sensors record device-dependent red--green--blue (RGB)
responses that must be mapped to a device-independent reference color
representation before subsequent rendering
\cite{Ierley2014,FinlaysonMackiewiczHurlbert2015}. The appropriate
camera-to-reference mapping depends on scene illumination, and
multi-illuminant camera characterization represents this dependence using
transforms calibrated under different illuminants
\cite{KaraimerBrown2018}. In a dual-illuminant construction, transforms
measured under two calibration illuminants are interpolated for intermediate
white points \cite{KaraimerBrown2018}. The Digital Negative (DNG) specification supports one, two, or three color-calibration sets and specifies
white-balance-dependent interpolation when multiple sets are present
\cite{DNGSpec171}.

Karaimer and Brown showed that extending conventional two-illuminant
interpolation with an additional pre-calibrated illuminant can improve camera
color reproduction by increasing characterization capacity across
illumination conditions \cite{KaraimerBrown2018}. DNG likewise provides a
third calibration state and corresponding third ColorMatrix, ForwardMatrix,
and ProfileHueSatMap data \cite{DNGSpec171}. When two or three
Hue/Saturation/Value mapping tables are present, DNG interpolates them using
the same calibration interpolation mechanism as the color-calibration tags
\cite{DNGSpec171}. The matrix characterization and HueSatMap payloads
assigned to these states therefore share the same host-defined illuminant
geometry. Condition-dependent nonlinear representations are also established
outside DNG profiling. Rota et al.\ introduced illuminant-adaptive
three-dimensional lookup tables (3D LUTs) for nonlinear camera color
correction \cite{C2LUT}, while Zeng et al.\ combined multiple basis 3D LUTs
using content-dependent weights for image-adaptive photo enhancement
\cite{ZengAdaptive3DLUT}.

We consider a setting in which an existing dual-illuminant characterization
is treated as a protected camera mapping while additional
condition-dependent nonlinear rendering capacity is required within the same
profile runtime. A conventional three-state DNG profile introduces an
additional characterization state together with the third HueSatMap payload,
and previous multi-illuminant camera-characterization methods similarly use
additional calibration states to increase characterization capacity
\cite{DNGSpec171,KaraimerBrown2018}. The problem considered here is to use
the additional shared interpolation degree of freedom for nonlinear rendering
while constraining the executed camera characterization to remain within a
prescribed deviation from the protected dual mapping. To the best of our knowledge, prior multi-illuminant camera-profile
constructions have not used an additional shared calibration slot whose
pre-serialization ColorMatrix and ForwardMatrix states are constructed from
the original dual matrix segments while the same host-defined interpolation
geometry carries an additional condition-dependent HueSatMap rendering basis
\cite{DNGSpec171,KaraimerBrown2018}.

We introduce \emph{Constrained Color Carrier} (CCC) for this setting. CCC
constructs the three pre-serialization ColorMatrix and ForwardMatrix states
from the original dual matrix segments, while the corresponding HueSatMap
payloads provide three nonlinear rendering bases. Condition-specific
HueSatMap payloads are solved as numerical preimages of the desired rendering
through the fixed serialized downstream operator, and the resulting payload
family is approximated jointly by the three carrier bases. Carrier selection
enforces a preservation bound evaluated on the host-effective interpolated
characterization. CCC therefore imposes the matrix-segment constraint and
host-effective characterization preservation as separate requirements.

The representation is evaluated through a controlled comparison of the
Dual, synthetic Ordinary Triple, Affine-only Triple, and CCC
representations, varying rendering-basis capacity, characterization geometry,
and carrier-selection criterion. The experiments use the Sony ILCE-7RM5
Adobe Standard dual-illuminant profile as the protected characterization
and a recovered white-balance-dependent Standard ColorCorrect rendering
from Phocus 4.0.1 for the Hasselblad X2D 100C as a real-world conditional
target. The final serialized DNG representation is verified by payload
readback and comparison between artifact-readback and solver-side offline
execution.
\section{Method}
\label{sec:method}

\subsection{Shared Multi-Illuminant Profile Representation}
\label{sec:shared_representation}

In the DNG representation considered here, camera
characterization and nonlinear rendering can share the same
white-balance-dependent calibration geometry \cite{DNGSpec171}. A
dual-illuminant profile can contain two ColorMatrix (CM) and ForwardMatrix
(FM) calibration sets and two corresponding Hue/Saturation/Value mapping
tables, hereafter referred to as HueSatMaps (HSMs). We denote the matrix
endpoints by $(CM_A,FM_A)$ and $(CM_D,FM_D)$, and the associated HSM
payloads by $B_A$ and $B_D$. For the profile subclass considered in this
work, the protected calibration endpoints are illuminant A and D65; the
compact subscript $D$ below denotes the D65 endpoint.

Throughout this work, $w_i(xy;\theta)$ denotes the normalized interpolation
weight assigned by the evaluated three-calibration host model to slot $i$ at
white-point chromaticity $xy$. The carrier parameters $\theta$ include the
three calibration descriptors that define this interpolation geometry. The
weights are determined by the host interpolation rule rather than optimized
as independent variables, and satisfy
\begin{equation}
w_i(xy;\theta)\geq0,
\qquad
\sum_{i=1}^{3}w_i(xy;\theta)=1.
\label{eq:host_weight_normalization}
\end{equation}
The same calibration weights are used for the matrix states and, when
multiple HSM tables are present, for the HSM payloads
\cite{DNGSpec171}. The concrete three-calibration weight evaluation used in
this work follows the public Adobe DNG SDK implementation, using the current
white point and the three stored calibration descriptors
\cite{AdobeDNGSDK171}. For compact notation,
$w_i(q;\theta)$ denotes these weights for requested condition $q$; the
host-effective white-point dependence is made explicit in
Sec.~\ref{sec:host_effective}.

This shared representation becomes restrictive when the protected
characterization is to retain its original dual family while the rendering
branch requires three HSM bases. A conventional three-slot profile supplies
\begin{equation}
(C_1,B_1),\qquad
(C_2,B_2),\qquad
(C_3,B_3),
\label{eq:shared_triple}
\end{equation}
so the third rendering payload is accompanied by a third characterization
state in the same calibration structure \cite{DNGSpec171}.

\begin{figure}[!htbp]
    \centering
    \includegraphics[width=\linewidth]
    {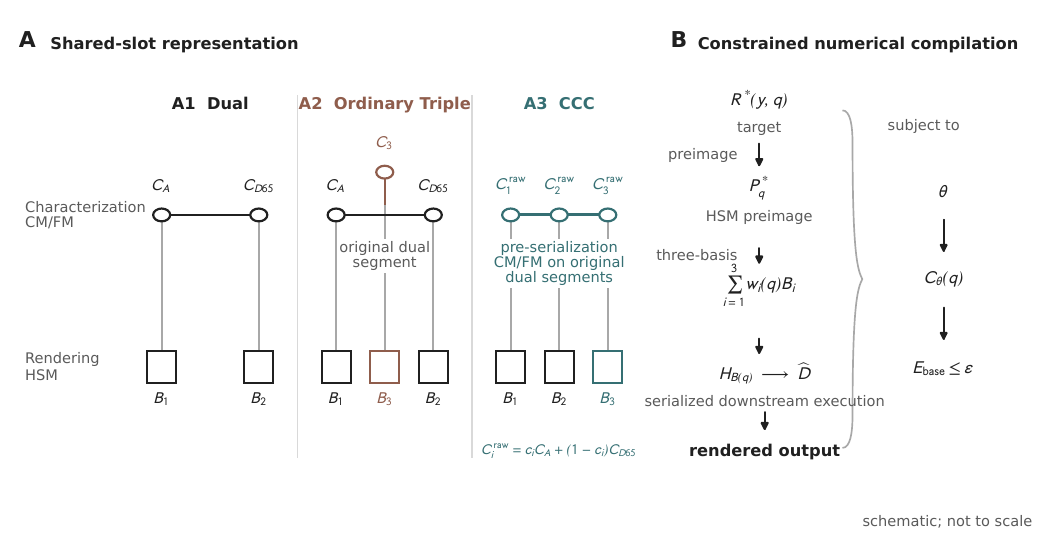}
    \caption{
Overview of the shared-slot representation and CCC compilation.
(A) A Dual profile provides two characterization states and two HueSatMap
payloads. An Ordinary Triple adds a third characterization state together
with a third HueSatMap payload, whereas CCC constructs the three
pre-serialization ColorMatrix/ForwardMatrix states from the original dual
matrix segments while using three HSM rendering bases.
(B) Condition-specific HSM preimages of the target rendering are
approximated by the three carrier bases and executed through the serialized
downstream operator, subject to the host-effective
characterization-preservation constraint.
The representation geometry is schematic and not to scale.
}
    \label{fig:ccc_construction}
\end{figure}

Let $\theta$ denote the carrier parameters and
$B=\{B_i\}_{i=1}^{3}$ the rendering bases. We formulate the representation objective as
\begin{equation}
\min_{\theta,B}
E_{\mathrm{target}}(\theta,B)
\label{eq:main_objective}
\end{equation}
subject to
\begin{equation}
E_{\mathrm{base}}(\theta)\leq\epsilon,
\qquad
\theta\in\Theta_{\mathrm{carrier}},
\label{eq:main_constraint}
\end{equation}
where $E_{\mathrm{target}}$ measures conditional-rendering error and
$E_{\mathrm{base}}$ measures deviation from the protected characterization.
The experimental definitions of these quantities are given in
Sec.~\ref{sec:evaluation}. Figure~\ref{fig:ccc_construction} summarizes the
shared-slot representation and its constrained numerical compilation.

\FloatBarrier

\subsection{Constrained Carrier Construction}
\label{sec:carrier_construction}

CCC uses the three-slot DNG representation to carry
three HSM rendering bases while constructing the corresponding
pre-serialization ColorMatrix and ForwardMatrix states from the original dual
matrix families. Let $CM_A$, $CM_D$, $FM_A$, and $FM_D$ denote the
protected dual-illuminant matrix endpoints. The pre-serialization carrier
matrices are constructed as
\begin{equation}
CM_i^{\mathrm{raw}}
=
c_iCM_A+(1-c_i)CM_D,
\qquad i=1,2,3,
\label{eq:ccc_cm}
\end{equation}
and
\begin{equation}
FM_i^{\mathrm{raw}}
=
c_iFM_A+(1-c_i)FM_D,
\qquad i=1,2,3,
\label{eq:ccc_fm}
\end{equation}
with
\begin{equation}
0\leq c_i\leq1.
\label{eq:ccc_coeff_bound}
\end{equation}
The ColorMatrix and ForwardMatrix within a slot use the same coefficient
$c_i$, placing the two matrices at corresponding positions along their
respective protected dual segments.

Before host evaluation, the raw carrier matrices are converted to their
serialized and runtime-interpreted forms using the same normalization and
quantization rules as the evaluated implementation:
\begin{equation}
CM_i^{s}
=
Q_{10^{-4}}\!\left(N_C(CM_i^{\mathrm{raw}})\right),
\qquad
FM_i^{s}
=
Q_{10^{-4}}\!\left(FM_i^{\mathrm{raw}}\right),
\label{eq:carrier_storage}
\end{equation}
followed by
\begin{equation}
FM_i^{r}
=
N_F(FM_i^{s}),
\label{eq:fm_runtime}
\end{equation}
where $Q_{10^{-4}}$ denotes element-wise quantization to four decimal
places, $N_C$ is the ColorMatrix white-normalization operator used by the
profile implementation, and
\begin{equation}
N_F(M)
=
\operatorname{diag}
\left(
\frac{XYZ_{D50}}{M\mathbf{1}}
\right)M
\label{eq:fm_normalization}
\end{equation}
is the runtime ForwardMatrix normalization. The serialized ColorMatrix
states $CM_i^{s}$ and runtime-interpreted ForwardMatrix states $FM_i^{r}$
are used in the host-effective characterization evaluation below.

Each rendering slot carries an HSM basis payload $B_i$. For condition $q$,
the host interpolation weights give
\begin{equation}
B(q;\theta)
=
\sum_{i=1}^{3}
w_i(q;\theta)B_i.
\label{eq:hsm_interpolation}
\end{equation}
The pre-serialization matrix branch is therefore restricted to the protected
dual families, while $B_1$, $B_2$, and $B_3$ provide three distinct nonlinear
rendering bases. The three basis payloads are fitted jointly in
Sec.~\ref{sec:numerical_solution}.

The carrier states and rendering bases use the native DNG
condition-descriptor fields and interpolation mechanism
\cite{DNGSpec171}. The
pre-serialization segment constraint in Eqs.~\eqref{eq:ccc_cm}--\eqref{eq:ccc_fm}
does not by itself determine the characterization executed by the host.
Storage normalization and quantization alter the serialized matrix states,
while the interpolation weights also depend on the candidate profile's
white-point solution. The executed mapping is therefore evaluated explicitly
below.

\subsection{Host-Effective Characterization Preservation}
\label{sec:host_effective}

The DNG conversion from CameraNeutral to white-point chromaticity is
iterative, and the recovered white point determines the calibration
interpolation \cite{DNGSpec171}. The weights defined in
Eq.~\eqref{eq:host_weight_normalization} must therefore be recomputed along
the candidate profile's white-point trajectory rather than evaluated once at
a nominal color temperature. Characterization preservation is evaluated
using the serialized matrix states together with this host-effective
interpolation trajectory.

For a requested condition $q$, let $\mathbf{n}(q)$ denote the CameraNeutral
obtained from the protected characterization. Given a current
Commission Internationale de l'Eclairage (CIE) $xy$ chromaticity estimate
$xy^{(k)}$, the candidate profile evaluates
$w_i(xy^{(k)};\theta)$ and interpolates its serialized ColorMatrices as
\begin{equation}
CM^{(k)}
=
\sum_{i=1}^{3}
w_i\!\left(
xy^{(k)};\theta
\right)
CM_i^{s},
\label{eq:iterated_cm}
\end{equation}
where $CM_i^{s}$ denotes the serialized ColorMatrix state of slot $i$.

Let $AB$ denote the AnalogBalance matrix and $CC^{(k)}$ the
CameraCalibration matrix evaluated with the same calibration weights. The
corresponding CIE XYZ-to-camera transform is
\begin{equation}
M_{\mathrm{XYZ}\rightarrow\mathrm{cam}}^{(k)}
=
AB\,CC^{(k)}\,CM^{(k)}.
\label{eq:xyz_to_camera}
\end{equation}
The updated tristimulus vector is
\begin{equation}
\mathbf{v}^{(k+1)}
=
\left(
M_{\mathrm{XYZ}\rightarrow\mathrm{cam}}^{(k)}
\right)^{-1}
\mathbf{n}(q),
\label{eq:neutral_inverse}
\end{equation}
with a pseudoinverse used when required by the dimensionality of the camera
color space \cite{DNGSpec171}. The next chromaticity estimate is
\begin{equation}
xy^{(k+1)}
=
\left(
\frac{v_X^{(k+1)}}{
v_X^{(k+1)}+v_Y^{(k+1)}+v_Z^{(k+1)}
},
\frac{v_Y^{(k+1)}}{
v_X^{(k+1)}+v_Y^{(k+1)}+v_Z^{(k+1)}
}
\right).
\label{eq:neutral_xy_update}
\end{equation}

At convergence, the recovered chromaticity
$\widehat{xy}(q;\theta)$ defines the host-effective interpolation weights
\begin{equation}
\widehat{w}_i(q;\theta)
=
w_i\!\left(
\widehat{xy}(q;\theta);\theta
\right).
\label{eq:effective_weights}
\end{equation}
The interpolation weights therefore depend on the white point recovered from
the candidate characterization itself.

The same weights determine the effective ColorMatrix and ForwardMatrix,
\begin{equation}
CM_{\theta}(q)
=
\sum_{i=1}^{3}
\widehat{w}_i(q;\theta)CM_i^{s},
\label{eq:effective_cm}
\end{equation}
and
\begin{equation}
FM_{\theta}(q)
=
\sum_{i=1}^{3}
\widehat{w}_i(q;\theta)FM_i^{r},
\label{eq:effective_fm}
\end{equation}
where $FM_i^{r}$ denotes the runtime-interpreted ForwardMatrix.

For profiles containing ForwardMatrix tags, the DNG camera-to-CIE-XYZ
transform referenced to standard illuminant D50 is
\begin{equation}
M_{\mathrm{cam}\rightarrow XYZ_{D50}}
=
FM_{\theta}(q)\,
D_{\theta}(q)\,
\left[
AB\,CC_{\theta}(q)
\right]^{-1},
\label{eq:camera_to_xyz_d50}
\end{equation}
where
\begin{equation}
\mathbf{n}_{\mathrm{ref}}(q;\theta)
=
\left[
AB\,CC_{\theta}(q)
\right]^{-1}
\mathbf{n}(q),
\qquad
D_{\theta}(q)
=
\operatorname{diag}
\left(
\mathbf{n}_{\mathrm{ref}}(q;\theta)
\right)^{-1}.
\label{eq:dng_reference_neutral}
\end{equation}
Equations~\eqref{eq:xyz_to_camera}--\eqref{eq:dng_reference_neutral}
follow the DNG camera-color model \cite{DNGSpec171}.

Carrier geometry can consequently alter the executed characterization
through the recovered white point, interpolation weights, serialized matrices,
and forward transform even when every pre-serialization ColorMatrix and
ForwardMatrix lies on its corresponding protected dual segment. We denote
the resulting characterization by $C_{\theta}(\mathbf{x},q)$ and the
protected mapping by $C_0(\mathbf{x},q)$. Preservation is imposed through
\begin{equation}
E_{\mathrm{base}}(\theta)
=
d_{\mathrm{base}}
\left(
C_{\theta},C_0
\right)
\leq\epsilon.
\label{eq:characterization_preservation}
\end{equation}
The output-domain realization of $d_{\mathrm{base}}$ compares the
protected reference path with the corresponding carrier-substituted
base-rendering path, with the added conditional target stage excluded.
The sampled domain, aggregation statistic, and numerical tolerance are
defined in Sec.~\ref{sec:evaluation}.

\subsection{Conditional Rendering Compilation}
\label{sec:rendering_compilation}

Let $\mathbf{y}$ denote the output of the protected characterization and
$R^{*}(\mathbf{y},q)$ the desired rendering for condition $q$. The
downstream representation used for deployment is fixed before the
condition-specific HSM payloads are solved; its serialized action is denoted
by $\widehat{\mathcal{D}}$.

Let $\mathcal{H}_{P}$ denote the HSM transform defined by payload $P$.
For each condition $q$, we solve
\begin{equation}
P_q^{*}
=
\arg\min_{P}
\mathcal{L}
\left(
\widehat{\mathcal{D}}
\left[
\mathcal{H}_{P}(\mathbf{y})
\right],
R^{*}(\mathbf{y},q)
\right),
\label{eq:preimage_payload}
\end{equation}
where $\mathcal{L}$ measures output error over the compilation sample set.
Numerically, Eq.~\eqref{eq:preimage_payload} is realized by a nodewise
inverse solve in RGB space. For an HSM source node $\mathbf{u}$, let
$\mathbf{r}_{q,e}(\mathbf{u})$ denote the corresponding target output at
exposure offset $e$. The intermediate RGB preimage is obtained from
\begin{equation}
\mathbf{z}_{q}^{*}(\mathbf{u})
=
\arg\min_{\mathbf{z}\in[0,4]^3}
\sum_{e=-3}^{3}
\left\|
\omega_e
\left[
\widehat{\mathcal{D}}\!\left(2^e\mathbf{z}\right)
-
\mathbf{r}_{q,e}(\mathbf{u})
\right]
\right\|_2^2,
\qquad
\omega_e=2^{-0.53e}.
\label{eq:nodewise_preimage}
\end{equation}
The solved RGB mapping
$\mathbf{u}\mapsto\mathbf{z}_{q}^{*}(\mathbf{u})$ is then converted to the
HueSatMap hue-shift, saturation-scale, and value-scale representation to form
$P_q^{*}$. The resulting $P_q^{*}$ is an independently solved,
condition-specific numerical preimage of the desired rendering through the
fixed downstream operator. We refer to the family $\{P_q^{*}\}$ as the
independent preimages. This formulation is an
instance of input precompensation through a fixed downstream system
\cite{WetzsteinBimber2007}; here the optimized input is an HSM payload
inside the camera-profile representation.

The three carrier bases jointly approximate the family
$\{P_q^{*}\}$ through the host interpolation in
Eq.~\eqref{eq:hsm_interpolation}. Let $B$ denote the stacked basis payloads.
At reference condition $q_0$, the interpolated payload is constrained by
\begin{equation}
\mathbf{a}^{\mathsf T}B
=
P_{q_0}^{*},
\qquad
a_i
=
\widehat{w}_i(q_0;\theta),
\label{eq:anchor_constraint}
\end{equation}
where the anchor weights are the host-effective weights defined in
Eq.~\eqref{eq:effective_weights}.

\subsection{Numerical Solution}
\label{sec:numerical_solution}

The carrier geometry contains a small nonlinear parameter block, whereas the
three HSM bases contain a much larger linear parameter block once the carrier
is fixed. The numerical procedure exploits this separation by updating the
carrier geometry in an outer search and eliminating the rendering-basis
coordinates through conditional linear solves.

For $K$ sampled carrier-control conditions, let
$W\in\mathbb{R}^{K\times3}$ contain the current host-effective interpolation
weights, with row $k$ given by
$\widehat{\mathbf{w}}^{\mathsf T}(q_k;\theta)$. Let
$\mathbf{g}\in\mathbb{R}^{K}$ denote the interpolation coordinate of the
protected dual characterization over the same conditions. For fixed slot
descriptors, the carrier coefficients are refreshed by the bounded
least-squares problem
\begin{equation}
\mathbf{c}^{*}
=
\arg\min_{\mathbf{c}}
\left\|
W\mathbf{c}-\mathbf{g}
\right\|_2^2,
\qquad
0\leq c_i\leq1.
\label{eq:carrier_ls}
\end{equation}
Because $W$ depends on the white point recovered from the candidate profile,
the coefficient update and the host white-point calculation are alternated
until the coefficient refresh converges. This step reduces the low-dimensional
carrier fit for fixed descriptors but does not analytically eliminate the
white-point coupling.

For fixed carrier geometry, let $\mathbf{a}^{\mathsf T}$ denote the anchor
row in Eq.~\eqref{eq:anchor_constraint}, and let $N$ span its null space.
The stacked rendering bases are parameterized as
\begin{equation}
B
=
B_0+NZ,
\label{eq:nullspace_parameterization}
\end{equation}
where $B_0$ satisfies
\begin{equation}
\mathbf{a}^{\mathsf T}B_0
=
P_{q_0}^{*}.
\label{eq:anchor_particular_solution}
\end{equation}
Let $W_f$ contain the interpolation weights for the non-anchor conditions and
let $P_f^{*}$ stack their condition-specific preimage payloads. Defining
\begin{equation}
Y_f
=
P_f^{*}-W_fB_0,
\label{eq:reduced_target}
\end{equation}
the remaining basis coordinates are obtained from
\begin{equation}
Z^{*}
=
\arg\min_Z
\left\|
S^{1/2}
\left(
W_fNZ-Y_f
\right)
\right\|_F^2,
\label{eq:varpro_basis}
\end{equation}
where $S$ contains the condition weights. This conditional elimination
follows the variable-projection principle for separable nonlinear
least-squares problems \cite{GolubPereyra1973,OLearyRust2013}. The
high-dimensional HSM basis coordinates are therefore solved conditionally for
each carrier state instead of being included directly in the outer nonlinear
search.

The payload least-squares problem controls aggregate approximation error, but
the reported rendering criterion is sensitive to high-error conditions. We
therefore repeat the basis solve with minimax-oriented reweighting: conditions
with larger rendered errors receive increased weight in subsequent solves, and
the evaluated iterate with the lowest worst-state rendering error is retained.
The reweighting schedule used in the experiments is reported in
Sec.~\ref{sec:experimental_setup}.

The numerical outer search generates and ranks carrier candidates on a
finite active condition set using the search score reported in
Sec.~\ref{sec:experimental_setup}. Each candidate is subsequently evaluated
using all 40,000 preservation samples at every active condition, and
candidates whose active-set preservation error exceeds $\epsilon$ are
rejected. Final acceptance is determined by the complete discretized-domain
audit described below.

Each accepted active-set candidate is finally audited over the complete
discretized preservation domain. When the audit identifies near-worst regions
not adequately represented by the active set, additional conditions from
those regions are inserted and the constrained solve is repeated. This
solve--audit--exchange procedure follows the general constraint-generation
principle used in semi-infinite optimization \cite{ZhangWuLopez2010}.
The exchange rule, numerical budget, and maximum number of cycles are
specified in Sec.~\ref{sec:experimental_setup}. Feasibility established by
this procedure applies to the evaluated discretization and is not a global
certificate over the continuous condition domain.
\section{Experiments}
\label{sec:experiments}

\subsection{Experimental Setup}
\label{sec:experimental_setup}

The experiments use the Sony ILCE-7RM5 Adobe Standard dual-illuminant
profile, whose original camera characterization defines the protected mapping
$C_0$. The evaluated profile is three-channel and uses identity AnalogBalance
and CameraCalibration terms. The conditional target $R^{*}$ is an executable
reconstruction of the white-balance-dependent Standard ColorCorrect stage of
Phocus 4.0.1 for the Hasselblad X2D 100C. The recovered model contains three $105\times89$ normalized two-channel
chroma fields anchored at 3100, 5550, and 9100~K, together with the
recovered temperature-dependent interpolation rule and the fixed color-domain
transforms required to execute the ColorCorrect stage. These recovered numerical components are fixed across
all profile representations evaluated in this study. The Film Tone stage is
excluded from $R^{*}$ and from the target-error evaluation. Independent
reverse engineering of the same Phocus processing stage has likewise reported
white-balance-dependent ColorCorrect behavior \cite{VLogAlchemy}.

The evaluation covers integer color temperatures from 2400 to 10000~K at
1-K intervals, giving 7,601 temperature states. For characterization
preservation, query tint is fixed at zero and a fixed set of 40,000
pseudorandom coordinates
$\mathbf{u}\sim U([0,1]^3)$ is generated once and reused at every
temperature. The corresponding camera-domain samples are
\begin{equation}
\mathbf{x}(T)
=
\mathbf{u}\odot
\mathrm{CameraWhite}_{0}(T),
\label{eq:preservation_samples}
\end{equation}
where $\mathrm{CameraWhite}_{0}(T)$ is obtained from the protected profile.
The pseudorandom seed is 20260826. Target rendering is evaluated at seven
exposure-value (EV) offsets,
\[
e\in\{-3,-2,-1,0,1,2,3\},
\]
giving 53,207 temperature--exposure states. Each state uses the same
1,458-color set comprising 729 pseudorandom samples drawn uniformly from
the RGB unit cube and a regular $9^3$ hue--saturation--value (HSV) grid.
This sample set uses pseudorandom seed 260237 and is reused across all
temperatures and exposure offsets. CCC compilation and target evaluation
operate on these samples in linear ProPhoto RGB.

The serialized CCC representation uses three $72\times32\times32$ HSM
bases and a $108\times32\times24$ Adobe Look table. The downstream
Creative Profile uses a $72\times16\times16$ lookup table. All formal
numerical comparisons use payloads read back from the serialized DNG Camera Profile (DCP) and Extensible Metadata Platform (XMP)
artifacts.

The nodewise preimage problem in
Eq.~\eqref{eq:nodewise_preimage} is solved using 10 damped local
least-squares iterations with a finite-difference step of
$2\times10^{-5}$, normal-equation damping of $2\times10^{-6}I$, and a
maximum per-iteration component step of $0.15$. The preimage RGB values are
constrained to $[0,4]^3$. The carrier search and subsequent conditional-rendering carrier refinement
use Powell's derivative-free method; the initial carrier search uses three
deterministic starts. The descriptor-temperature
bounds are
\[
T_1\in[1667,5000]\ {\rm K},\qquad
T_2\in[3000,8000]\ {\rm K},\qquad
T_3\in[6000,25000]\ {\rm K},
\]
with each descriptor tint constrained to $[-200,200]$ and each carrier
coefficient constrained to $[0,1]$. These intervals are numerical search
bounds rather than measured physical calibration illuminants. The carrier
and conditional-rendering refinement stages each use a budget of 120 outer
evaluations. During conditional-rendering refinement, candidates are ranked
using
\begin{equation}
S
=
E_{\mathrm{target}}^{\max}
+
0.15E_{\mathrm{target}}^{\mathrm{mean}}
+
100
\max
\left[
0,\,
E_{\mathrm{base}}^{\mathrm{sketch}}
-
(\epsilon-1.5\times10^{-4})
\right],
\label{eq:search_score}
\end{equation}
where the target metrics are defined in
Sec.~\ref{sec:evaluation} and
$E_{\mathrm{base}}^{\mathrm{sketch}}$ is evaluated using the first
4,096 preservation samples. Candidate ranking is followed by the full
40,000-sample active-set preservation test at the original tolerance
$\epsilon$. HSM basis refinement begins with equal temperature weights.
After each basis solve, the worst-exposure state-wise $P_{95}$ defined in
Sec.~\ref{sec:evaluation} is evaluated at every active temperature.
Temperatures at or above the 75th percentile of these errors receive twice
their previous weight, after which the weights are renormalized to unit mean.
Six reweighting updates produce seven solve--evaluation rounds, and the round
with the smallest maximum state-wise $P_{95}$ is retained.

The full-domain preservation audit permits at most four
solve--audit--exchange cycles. Exchange candidates are drawn from near-worst
regions whose preservation error reaches at least $0.98$ of the current
full-domain maximum, using a 50-K neighborhood and adding at most eight
temperatures per cycle. The complete 7,601-temperature audit is used as the
final preservation test for every reported carrier. The implementation of the profile-construction and evaluation pipeline,
including the three-calibration host-weight evaluation and the code used to
execute the recovered conditional target, is publicly available at
\url{https://github.com/fishvivfish/HNCS-Lightroom}. The recovered Phocus
numerical assets required by the target model are not redistributed.

\subsection{Evaluation}
\label{sec:evaluation}

We compare four profile representations. All four use the same protected
dual-illuminant profile, conditional target, serialized downstream
representation, sampling domains, and error definitions. The Dual control
retains the original two characterization states and uses two HSM rendering
bases fitted to the same condition-specific target payloads defined in
Sec.~2.4. For the synthetic Ordinary Triple control, the third calibration descriptor
was fixed before the controlled comparison at 5000~K with zero tint. The 5000-K location is motivated by the approximately 5000-K additional
interpolation control point considered by Karaimer and Brown
\cite{KaraimerBrown2018}; zero tint is a fixed control setting rather
than a measured third-illuminant parameter. Its ColorMatrix
and ForwardMatrix are obtained from the host-effective interpolation of the
protected dual characterization at 5000~K followed by the same storage
normalization used for the evaluated profiles. The representation carries
three HSM bases. It therefore represents a synthetic third-calibration control,
not an independently measured physical third-illuminant calibration.

The Affine-only Triple and CCC both construct their three pre-serialization
ColorMatrix and ForwardMatrix states from the original dual matrix segments. The Affine-only carrier is
selected by fitting the nominal interpolation trajectory and is subsequently
audited using the host-effective preservation metric. CCC evaluates candidate
carriers using the host-effective characterization and applies the preservation
criterion during carrier selection.

\begin{table}[t]
\centering
\caption{Profile representations used in the comparison.}
\label{tab:representations}
\small
\setlength{\tabcolsep}{4pt}
\begin{tabularx}{\linewidth}{@{}l L c L@{}}
\toprule
Profile &
Characterization states &
HSM bases &
Carrier criterion \\
\midrule
Dual &
Original dual &
2 &
--- \\

Ordinary Triple &
Synthetic third state &
3 &
--- \\

Affine-only Triple &
Dual-segment carrier &
3 &
Nominal trajectory \\

CCC &
Dual-segment carrier &
3 &
Host-effective preservation \\
\bottomrule
\end{tabularx}
\end{table}

Preservation and target-rendering fidelity are evaluated separately.
$E_{\mathrm{base}}$ measures deviation of the protected camera mapping after
carrier substitution with the new conditional target rendering excluded,
whereas $E_{\mathrm{target}}$ measures approximation of the recovered
conditional renderer through the evaluated downstream chain. Color errors
are measured as Euclidean distances in the Oklab perceptual color space
\cite{Ottosson2020}. Colors in the linear ProPhoto RGB D50 reference domain
are converted to CIE XYZ D50 and chromatically adapted to D65 using the
Bradford transform before conversion to Oklab. Distances are reported in the
native Oklab coordinate scale without multiplication by 100:
\begin{equation}
d_{\mathrm{OK}}
\left(
\mathbf{z},\mathbf{z}^{*}
\right)
=
\left\|
\operatorname{Oklab}(\mathbf{z})
-
\operatorname{Oklab}(\mathbf{z}^{*})
\right\|_2.
\label{eq:oklab_distance}
\end{equation}

At each preservation condition, the 40,000 sample-wise distances are
summarized by their 95th percentile. The preservation diagnostic compares
the protected reference path with the corresponding carrier-substituted path
after the new conditional target rendering is removed. The candidate path
includes the serialized re-representation of the original Adobe Look;
consequently, this diagnostic measures preservation of the complete evaluated
base-rendering path rather than a matrix-only ColorMatrix/ForwardMatrix
difference. Let $P_{95}^{\mathrm{base}}(T)$ denote this value at temperature
$T$. The preservation metric is
\begin{equation}
E_{\mathrm{base}}
=
\max_T
P_{95}^{\mathrm{base}}(T).
\label{eq:base_metric}
\end{equation}
The preservation criterion used throughout the study is
\begin{equation}
E_{\mathrm{base}}\leq\epsilon,
\qquad
\epsilon=0.004.
\label{eq:preservation_tolerance}
\end{equation}
The value of $\epsilon$ is an engineering feasibility tolerance for this
study and is not defined as a perceptual visibility threshold.

For the conditional target, let
$P_{95}^{\mathrm{target}}(T,e)$ denote the 95th-percentile error over the
1,458-color sample set at temperature $T$ and exposure offset $e$. We report
\begin{equation}
E_{\mathrm{target}}^{\max}
=
\max_{T,e}
P_{95}^{\mathrm{target}}(T,e),
\qquad
E_{\mathrm{target}}^{\mathrm{mean}}
=
\frac{1}{N}
\sum_{T,e}
P_{95}^{\mathrm{target}}(T,e),
\label{eq:target_metrics}
\end{equation}
where $N=53{,}207$. The first quantity is the worst-condition state-wise
95th-percentile error, and the second is the mean of the state-wise
95th percentiles over the complete evaluation domain.

The residual-error diagnostic uses 26 temperatures spanning
2400--10000~K, including the 5550-K reference condition. The target-only
carrier is optimized for target-rendering fidelity without the
characterization-preservation constraint. Under both the target-only carrier
and the preservation-constrained CCC carrier, we evaluate the independent
preimages and their three-basis approximations using the same seven exposure
offsets and target-error definition.

\section{Results}
\label{sec:results}

\subsection{Representation Comparison}
\label{sec:representation_results}

\begin{figure}[!htbp]
    \centering
    \includegraphics[width=\linewidth]
    {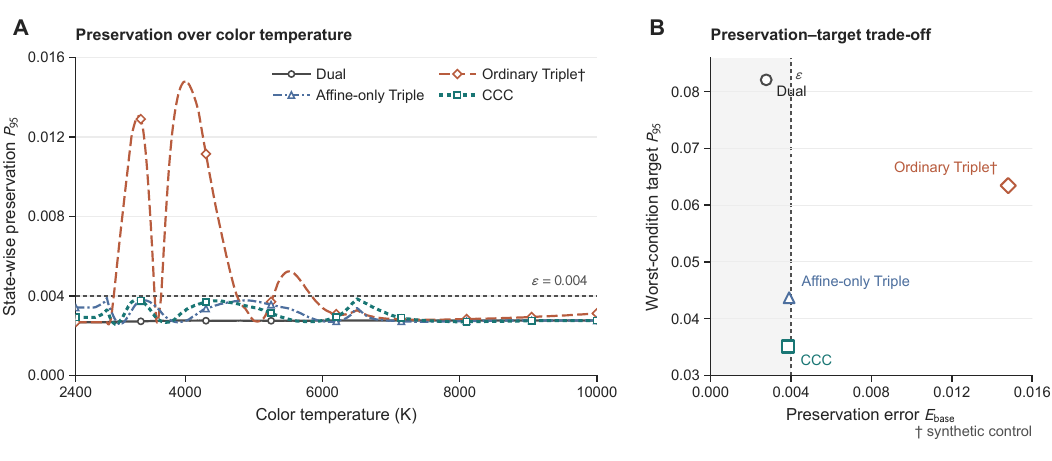}
    \caption{
    Representation comparison under the characterization-preservation
    constraint.
    (A) State-wise preservation $P_{95}$ over color temperature for the Dual,
    synthetic Ordinary Triple, Affine-only Triple, and CCC representations.
    The horizontal dashed line marks the preservation tolerance
    $\epsilon=0.004$.
    (B) Preservation error $E_{\mathrm{base}}$ versus worst-condition target
    $P_{95}$ for the same four representations. The vertical dashed line marks
    the preservation boundary; representations to its left satisfy the
    prescribed criterion. The Ordinary Triple is a synthetic
    third-calibration control.
    }
    \label{fig:representation_comparison}
\end{figure}

Figure~\ref{fig:representation_comparison} shows the preservation behavior of
the four profile representations over color temperature and their positions in
the preservation--target error plane. The Dual representation remains within
the preservation tolerance over the complete evaluated temperature range.
The Ordinary Triple exceeds the tolerance over a broad
intermediate-temperature region, whereas both dual-segment-constrained
three-basis representations remain within the prescribed tolerance.

\begin{table}[t]
\centering
\caption{Full-domain preservation and target-rendering errors for the four
profile representations.}
\label{tab:representation_results}
\small
\setlength{\tabcolsep}{3.5pt}
\begin{tabularx}{\linewidth}{@{}l c c c L@{}}
\toprule
Profile &
$E_{\mathrm{base}}$ &
$E_{\mathrm{target}}^{\mathrm{mean}}$ &
$E_{\mathrm{target}}^{\mathrm{max}}$ &
Worst target condition \\
\midrule
Dual &
0.002766 &
0.02314 &
0.08208 &
10000 K / $+3$ EV \\

Ordinary Triple &
0.014802 &
0.03027 &
0.06344 &
4006 K / $+3$ EV \\

Affine-only Triple &
0.003905 &
0.01984 &
0.04368 &
10000 K / $+3$ EV \\

CCC &
0.003857 &
0.01904 &
0.03508 &
4226 K / $+3$ EV \\
\bottomrule
\end{tabularx}
\end{table}

\FloatBarrier

The Dual representation gives the lowest preservation error,
$E_{\mathrm{base}}=0.002766$, but the largest worst-condition target error,
$E_{\mathrm{target}}^{\max}=0.08208$. The nonzero Dual preservation value is dominated by the serialized
re-representation of the original Adobe Look in the candidate path rather
than by a change in the Dual ColorMatrix or ForwardMatrix states. The Ordinary Triple lowers
$E_{\mathrm{target}}^{\max}$ to $0.06344$, while its preservation error
increases to $0.014802$ and exceeds the $0.004$ criterion. Its mean target
error is also higher than that of the Dual representation.

The Affine-only Triple satisfies the preservation criterion and lowers both
target-error metrics relative to the Dual representation. CCC also satisfies
the criterion, with $E_{\mathrm{base}}=0.003857$, and gives the lowest mean
and worst-condition target errors among the three-basis representations:
$E_{\mathrm{target}}^{\mathrm{mean}}=0.01904$ and
$E_{\mathrm{target}}^{\max}=0.03508$. Relative to the Dual representation,
CCC reduces the worst-condition metric by $57.3\%$. The reported CCC
preservation value is obtained from the complete discretized-domain audit;
no exchange update was triggered for the final solution.

\FloatBarrier

\subsection{Error and Capacity Analysis}
\label{sec:error_capacity_results}

\begin{figure}[!htbp]
    \centering
    \includegraphics[width=\linewidth]
    {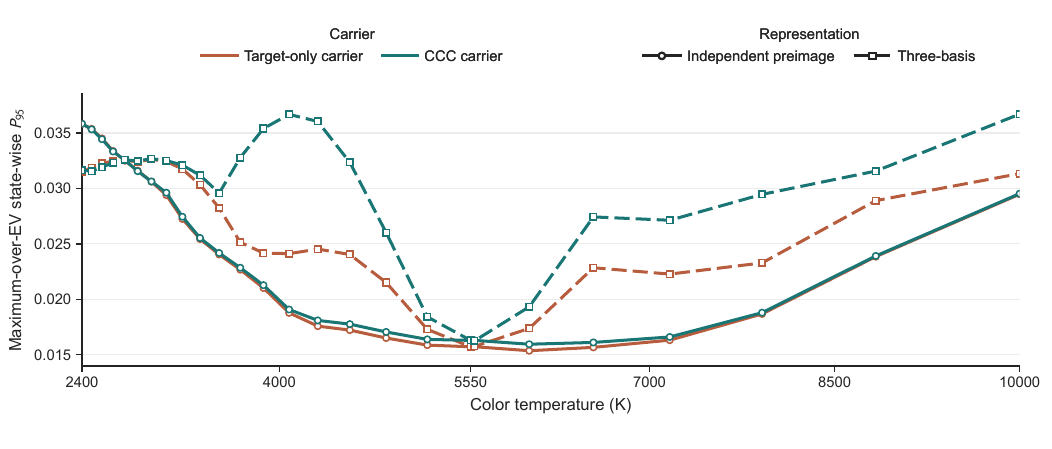}
    \caption{
    Residual target-error diagnostic over the 26-temperature evaluation set.
    For each temperature, the plotted value is the maximum over the seven
    exposure states of the state-wise target $P_{95}$.
    Color distinguishes the target-only and CCC carriers, while line style
    distinguishes independent preimages from their three-basis
    approximations.
    }
    \label{fig:error_capacity}
\end{figure}

Figure~\ref{fig:error_capacity} compares the independent preimages with their
three-basis approximations under the target-only and CCC carriers over the
26-temperature diagnostic set. The four curves remain of comparable
magnitude across the evaluated temperature range, while their relative
ordering varies with temperature and carrier setting.

\begin{table}[t]
\centering
\caption{Mean and maximum state-wise P95 target-rendering errors on the
26-temperature diagnostic set.}
\label{tab:diagnostic_results}
\small
\setlength{\tabcolsep}{3.5pt}

\begin{tabularx}{\linewidth}{
@{}
l
>{\raggedright\arraybackslash}X
c
c
l
@{}
}
\toprule
Carrier &
Representation &
Mean P95 &
Max P95 &
Worst condition \\
\midrule

Target-only &
Independent preimages &
0.01650 &
0.03582 &
2400 K / $-3$ EV \\

Target-only &
Three-basis &
0.01815 &
0.03263 &
2963 K / $-3$ EV \\

CCC &
Independent preimages &
0.01666 &
0.03584 &
2400 K / $-3$ EV \\

CCC &
Three-basis &
0.02056 &
0.03671 &
10000 K / $+3$ EV \\

\bottomrule
\end{tabularx}
\end{table}

Under the target-only carrier, the three-basis representation has a higher
mean state-wise $P_{95}$ but a lower maximum state-wise $P_{95}$ than the
independent preimages. Under CCC, the three-basis representation has a
higher mean, while the maximum changes only slightly upward. The reported
maximum metric therefore does not exhibit a common ordering between the two
representation levels across the two carrier settings.

Removing the characterization-preservation constraint improves the
full-domain target fit. Relative to CCC, the target-only carrier lowers
$E_{\mathrm{target}}^{\mathrm{mean}}$ from $0.01904$ to $0.01716$ and
$E_{\mathrm{target}}^{\max}$ from $0.03508$ to $0.03083$. Its preservation
error is $E_{\mathrm{base}}=0.011854$, exceeding the
$\epsilon=0.004$ preservation criterion.

\FloatBarrier

\subsection{Serialization and Offline Execution Validation}
\label{sec:serialized_validation}

The final profile was serialized and read back before numerical validation.
All evaluated color-table payloads are reproduced without numerical change,
as summarized in Table~\ref{tab:serialization}.

\begin{table}[t]
\centering
\caption{Serialized profile dimensions and payload-readback differences.}
\label{tab:serialization}
\small
\setlength{\tabcolsep}{4pt}
\begin{tabularx}{\linewidth}{@{}L c C@{}}
\toprule
Component &
Serialized dimensions &
\shortstack{Maximum absolute\\payload difference} \\
\midrule
Adobe Look &
$108\times32\times24$ &
0 \\

CCC HSM bases &
$3\times(72\times32\times32)$ &
0 \\

Downstream XMP lookup table &
$72\times16\times16$ &
0 \\
\bottomrule
\end{tabularx}
\end{table}

Execution using the artifact-readback payloads also yields zero numerical
difference from the corresponding solver-side offline execution. At the
5550-K reference condition, the interpolated serialized HSM payload differs from the prescribed anchor payload by at most
$1.91\times10^{-6}$.

\FloatBarrier
\section{Discussion}
\label{sec:discussion}

Conventional multi-illuminant camera characterization uses additional
calibration states to improve colorimetric mapping across illumination
conditions \cite{KaraimerBrown2018}, and the DNG specification places a
third calibration state within the same multi-calibration interpolation
structure \cite{DNGSpec171}. The representation comparison considers a
different use of this shared slot. The Ordinary Triple exceeds the
preservation criterion, whereas the Affine-only Triple and CCC, whose
pre-serialization ColorMatrix and ForwardMatrix states are constructed from
the original dual matrix segments, both remain within the prescribed
tolerance. CCC also provides the lowest mean and worst-condition target
errors among the three-basis representations. Since the Affine-only Triple
also satisfies the preservation criterion, the comparison supports the role
of the dual-segment constraint relative to the Ordinary Triple construction
but does not show a separate preservation advantage from host-effective
carrier selection. In the evaluated representation, the additional shared
slot therefore increases nonlinear rendering capacity while the
pre-serialization matrix states remain within the protected dual families and
the executed characterization remains within the prescribed tolerance.

The independent preimages and their three-basis approximations have
comparable target errors over the diagnostic temperature set, and their
maximum state-wise $P_{95}$ values do not have a common ordering across the
two carrier settings. The two representation levels are optimized separately
under objectives that differ from the reported maximum state-wise $P_{95}$,
so their maxima need not be ordered. The independent-preimage result
therefore serves as a diagnostic reference before finite-basis approximation
rather than a lower bound on the three-basis metric. The experiment does not
identify finite-basis approximation as the dominant source of residual target
error. Removing the characterization-preservation constraint reduces both
full-domain target-error metrics but increases
$E_{\mathrm{base}}$ to $0.011854$, above the
$\epsilon=0.004$ criterion. The evaluated profile thus exhibits a trade-off
between conditional-rendering fidelity and characterization preservation.

Serialization and readback reproduce the evaluated color-table payloads
exactly, and execution using the artifact-readback payloads is numerically
identical to the corresponding solver-side offline execution. The final CCC
representation therefore survives the evaluated serialization path without
numerical change. Agreement with execution in a commercial Adobe host remains
untested. The complete discretized-domain audit verifies preservation outside
the finite optimization set for the reported carrier. No exchange update was triggered for the final Sony ILCE-7RM5 solution, so
the exchange mechanism remains a numerical safeguard rather than an
empirically demonstrated requirement in the present experiment.

The current evaluation uses one protected dual-illuminant camera profile and
one recovered conditional-rendering family, so transfer across cameras and
targets remains untested. The target covers the white-balance-dependent
Standard ColorCorrect rendering recovered from Phocus 4.0.1 for the
Hasselblad X2D 100C and excludes the Film Tone stage. The Ordinary Triple is
a synthetic third-calibration control rather than an independently measured
physical third-illuminant calibration. The preservation audit varies color
temperature at zero tint and therefore samples only a one-dimensional subset
of the white-point domain. The tolerance $\epsilon=0.004$ is an engineering
feasibility criterion for this study and does not define a general perceptual
threshold, and feasibility over the complete discretized temperature domain
does not constitute a continuous-domain certificate. Direct commercial-host validation is required to establish agreement between
the evaluated offline execution model and deployed host execution.
Evaluation on additional protected camera profiles, broader two-dimensional
white-point domains, and other conditional-rendering targets is required to
assess transfer beyond the present setting.
\section{Conclusion}
\label{sec:conclusion}

This work introduced CCC, a shared-slot camera-profile construction that increases conditional nonlinear rendering
capacity while constructing its pre-serialization characterization matrices
from a protected dual-illuminant family. For the evaluated Sony ILCE-7RM5
Adobe Standard profile, CCC achieves a preservation error of
$E_{\mathrm{base}}=0.003857$, below the prescribed tolerance of
$\epsilon=0.004$, and reduces the worst-condition target $P_{95}$ from
$0.08208$ for the Dual representation to $0.03508$. The synthetic Ordinary
Triple reduces the worst-condition target error but violates the preservation
criterion, whereas the dual-segment-constrained three-basis representations
remain feasible. Serialized payload readback is exact, and artifact-readback
offline execution is numerically identical to solver-side execution. These
results show that, in the evaluated profile, additional conditional-rendering
capacity can be carried by the shared multi-illuminant representation while
keeping the pre-serialization ColorMatrix and ForwardMatrix states within the
protected dual matrix families and the executed characterization within the
prescribed tolerance.

\bibliographystyle{unsrt}
\bibliography{references}

@manual{DNGSpec171,
  author = {{Adobe Inc.}},
  title  = {Digital Negative ({DNG}) Specification, Version 1.7.1.0},
  year   = {2023},
  month  = sep,
  note   = {\url{https://helpx.adobe.com/camera-raw/desktop/dng-and-file-formats/digital-negative.html}}
}

@article{Ierley2014,
  author  = {Ierley, G. R.},
  title   = {On Accurate Color Calibration for Digital Cameras},
  journal = {Journal of Imaging Science and Technology},
  volume  = {58},
  number  = {1},
  pages   = {010501-1--010501-24},
  year    = {2014},
  doi     = {10.2352/J.ImagingSci.Technol.2014.58.1.010501}
}

@article{FinlaysonMackiewiczHurlbert2015,
  author  = {Finlayson, Graham D. and Mackiewicz, Michal and Hurlbert, Anya},
  title   = {Color Correction Using Root-Polynomial Regression},
  journal = {IEEE Transactions on Image Processing},
  volume  = {24},
  number  = {5},
  pages   = {1460--1470},
  year    = {2015},
  doi     = {10.1109/TIP.2015.2405336}
}

@inproceedings{KaraimerBrown2018,
  author    = {Karaimer, Hakki Can and Brown, Michael S.},
  title     = {Improving Color Reproduction Accuracy on Cameras},
  booktitle = {Proceedings of the IEEE Conference on Computer Vision and Pattern Recognition},
  pages     = {6440--6449},
  year      = {2018},
  month     = jun,
  doi       = {10.1109/CVPR.2018.00674}
}

@article{ZengAdaptive3DLUT,
  author  = {Zeng, Hui and Cai, Jianrui and Li, Lida and Cao, Zisheng and Zhang, Lei},
  title   = {Learning Image-Adaptive {3D} Lookup Tables for High Performance Photo Enhancement in Real-Time},
  journal = {IEEE Transactions on Pattern Analysis and Machine Intelligence},
  volume  = {44},
  number  = {4},
  pages   = {2058--2073},
  year    = {2022},
  doi     = {10.1109/TPAMI.2020.3026740}
}

@misc{C2LUT,
  author        = {Rota, Claudio and Cogo, Luca and Bianco, Simone and Schettini, Raimondo},
  title         = {Illuminant-Adaptive {3D} Lookup Tables for Camera Color Correction},
  year          = {2026},
  eprint        = {2607.11681},
  archivePrefix = {arXiv},
  primaryClass  = {cs.CV},
  doi           = {10.48550/arXiv.2607.11681},
  note          = {\url{https://arxiv.org/abs/2607.11681}}
}

@article{GolubPereyra1973,
  author  = {Golub, G. H. and Pereyra, V.},
  title   = {The Differentiation of Pseudo-Inverses and Nonlinear Least Squares Problems Whose Variables Separate},
  journal = {SIAM Journal on Numerical Analysis},
  volume  = {10},
  number  = {2},
  pages   = {413--432},
  year    = {1973},
  doi     = {10.1137/0710036}
}

@article{OLearyRust2013,
  author  = {O'Leary, Dianne P. and Rust, Bert W.},
  title   = {Variable Projection for Nonlinear Least Squares Problems},
  journal = {Computational Optimization and Applications},
  volume  = {54},
  number  = {3},
  pages   = {579--593},
  year    = {2013},
  doi     = {10.1007/s10589-012-9492-9}
}

@inproceedings{WetzsteinBimber2007,
  author    = {Wetzstein, Gordon and Bimber, Oliver},
  title     = {Radiometric Compensation through Inverse Light Transport},
  booktitle = {15th Pacific Conference on Computer Graphics and Applications},
  pages     = {391--399},
  year      = {2007},
  doi       = {10.1109/PG.2007.47}
}

@article{ZhangWuLopez2010,
  author  = {Zhang, Liping and Wu, Soon-Yi and L{\'o}pez, Marco A.},
  title   = {A New Exchange Method for Convex Semi-Infinite Programming},
  journal = {SIAM Journal on Optimization},
  volume  = {20},
  number  = {6},
  pages   = {2959--2977},
  year    = {2010},
  doi     = {10.1137/090767133}
}

@misc{Ottosson2020,
  author       = {Ottosson, Bj{\"o}rn},
  title        = {A Perceptual Color Space for Image Processing},
  year         = {2020},
  month        = dec,
  howpublished = {\url{https://bottosson.github.io/posts/oklab/}}
}

@misc{VLogAlchemy,
  author       = {{shenmintao}},
  title        = {{V-Log Alchemy (Lumix Body Snatcher)}},
  year         = {2026},
  howpublished = {GitHub repository},
  note         = {Hasselblad/Phocus reconstruction materials. \url{https://github.com/shenmintao/V-Log-Alchemy}}
}

@manual{AdobeDNGSDK171,
  author = {{Adobe Inc.}},
  title  = {Adobe Digital Negative ({DNG}) Software Development Kit},
  year   = {2026},
  month  = jun,
  note   = {Version 1.7.1, Build 2611, June 9, 2026. \url{https://helpx.adobe.com/camera-raw/desktop/dng-and-file-formats/digital-negative.html}}
}

\end{document}